\documentclass[12pt]{article}

\usepackage[numbers,compress]{natbib}
\usepackage{bm,bbm}
\usepackage{amsmath}
\usepackage{amssymb}
\usepackage{siunitx}
\usepackage{xcolor}
\usepackage{soul}
\usepackage{hyperref}
\usepackage{url}

\DeclareMathOperator{\tr}{tr}

\title{Dual Approaches to Express the Generalized Degree of Polarimetric Purity }

\author{A.~Bhattacharya, S.~Dey, A.~C.~Frery}
\date{}

\begin{document}

\maketitle

\section{Introduction}

The \textit{degree of polarimetric purity} is an invariant dimensionless quantity that characterizes the closeness of a polarization state of a wave to a pure state and is related to the Von Neumann entropy~\cite{brosseau1998fundamentals}. The polarimetric purity of a plane wave characterized by the second-order statistics (i.e., the covariance matrix) is uniquely described by the \textit{degree of polarization}. However, the 2D formalism is only applicable when the wave propagation direction is fixed. This assumption is typical in optical and radar polarimetric measurements. Therefore, one must consider all the components to describe the general state of wave polarization. Starting from Samson~\cite{samson1973}, and Barakat~\cite{barakat1977degree}, several different concepts have been proposed in the literature to describe the 3D \textit{degree of polarization}~\cite{ellis2005degree,refregier2006invariant,dennis2007three,setala2002degree,luis2005degree,gil2018polarimetric}.  

The \textit{generalized degree of polarimetric purity}, $P_{nD}$~\cite{san2011invariant,gil2004generalized,gil2007polarimetric} for the $n\times n$ Hermitian and positive semi-definite coherency matrix $\mathbf{\Phi}$ is defined as,
\begin{equation}
    P_{nD} = \left\{\frac{1}{n-1} \Bigg[\frac{n[\tr{(\mathbf{\Phi}^2)}]}{(\tr{\mathbf{\Phi}})^2} - 1 \Bigg] \right\}^{1/2}
    \label{eqn:purity_gil}
\end{equation}
where $\tr(\mathbf{\Phi})$ is the trace of $\mathbf{\Phi}$. This is an invariant dimensionless quantity satisfying, $0 \le P_{nD} \le 1$. The minimum value of the degree of purity $P_{nD} = 0$ corresponds to a completely random polarization state of a wave. In contrast, the maximum value $P_{nD} = 1$ corresponds to a completely pure polarization state of a wave. In between the two extremities, the wave is partially pure polarized. 

In this work, we propose two distinct approaches to express the \textit{generalized degree of polarimetric purity}, $P_{nD}$. In the first approach, we utilize the definition of the mean and standard deviation of real positive eigenvalues of Hermitian positive semi-definite matrices~\cite{bhattacharya2022scattering}. In the second approach, we use elementary concepts from vector calculus and align them with a matrix decomposition procedure following certain notions from Lie algebra~\cite{gilmore2012lie}. Finally, we establish the parity of the two approaches to compute the \textit{generalized degree of polarimetric purity}.

\section{Approach I: Coefficient of Variation}

The mean $(m)$ and the standard deviation $(s)$ of the real positive eigenvalues $\lambda_1 \geq \lambda_2 \geq \dots \geq \lambda_n \geq 0$ for a $n\times n$ coherency matrix $\mathbf{\Phi}$ are defined using a simple function of the trace of the matrix and the trace of its square given in~\cite{BoundsforEigenvaluesUsingTraces} as,
\begin{align}
m & = \frac{1}{n}\tr (\mathbf{\Phi}) = \frac{1}{n} \sum_{j=1}^n \lambda_j, \text{ and}	\\
s^2 & = \frac{1}{n} \Bigg[ \sum_{j=1}^{n}\lambda_j^2 - \frac1n \bigg(\sum_{j=1}^{n}{\lambda_j}\bigg)^2\Bigg] = 
\frac{\tr{(\mathbf{\Phi}^2)} - (\tr{\mathbf{\Phi}})^2/n}{n}, \\
& = \frac{\tr(\mathbf{\Phi}^2)}{n} - m^{2}.
\end{align}
Using these two quantities, we propose a new expression for the generalized degree of polarimetric purity as,
\begin{equation}
    P_{nD} = \frac{s}{\sqrt{n-1}\,m}
    \label{eqn:purity_bhattacharya}
\end{equation}
One can easily observe that for $n=2$, and $n=3$, the expressions for the generalized degree of polarimetric purity can be related as~\cite{bhattacharya2022scattering},  
\begin{align*}
    P_{2D} &= \left\{\frac{2[\tr{(\mathbf{\Phi}^2)}]}{(\tr{\mathbf{\Phi}})^2} - 1  \right\}^{1/2} = s/m \\
    P_{3D} &= \left\{\frac{1}{2} \Bigg[\frac{3[\tr{(\mathbf{\Phi}^2)}]}{(\tr{\mathbf{\Phi}})^2} - 1 \Bigg] \right\}^{1/2} = s/\sqrt{2}m 
\end{align*}
The proposed expression~\eqref{eqn:purity_bhattacharya} is physically intuitive as it directly relates the measure of polarimetric purity to the coefficient of variation (i.e., $s/m$) of the eigenvalues of a $n\times n$ matrix. 

Furthermore, let $\psi$ denote the angle between $\mathbf{\Phi}$ and the identity matrix $\mathbf{I}_{n}$ in the space of $n \times n$ matrices. Analogously, we can express this angle between the vector of eigenvalues $\lambda_{1}, \lambda_{2}, \dots, \lambda_{n}$ and the equiangular line as~\cite{BoundsforEigenvaluesUsingTraces},
\begin{equation}
    \psi = \cos^{-1}\left(\tr{\mathbf{\Phi}}/\sqrt{n\left(\tr{\mathbf{\Phi}}\right)^{2}}\right)
\end{equation}
therefore, using $\psi$ we can also express the generalized degree of polarimetric purity as,
\begin{equation}
    P_{nD} = \frac{\tan{\psi}}{\sqrt{n-1}}
\end{equation}
which immediately relates, that $\tan{\psi} = s/m$.

\section{Approach II: Direct Sum Decomposition}

Let $\mathbf{\Phi}$ be decomposed as,  
\begin{equation}
    \mathbf{\Phi} = \mathbf{\Phi}_{1} + \mathbf{\Phi}_{2} + \mathbf{\Phi}_{3}
\end{equation}
where,
\begin{align}
    \mathbf{\Phi}_{1} &= \dfrac{(\mathbf{\Phi} + \mathbf{\Phi}^{*})}{2} - \dfrac{\tr(\mathbf{\Phi})}{n}\mathbf{I}_{n}, \quad (\text{Traceless symmetric matrix})\\
    \mathbf{\Phi}_{2} &= \dfrac{(\mathbf{\Phi} - \mathbf{\Phi}^{*})}{2}, \quad (\text{Anti-symmetric matrix}) \\
    \mathbf{\Phi}_{3} &=  \dfrac{\tr(\mathbf{\Phi})}{n}\mathbf{I}_{n}, \quad (\text{Scalar matrix}) 
\end{align}
where $\mathbf{I}_{n}$ is the $n \times n$ identity matrix and $\mathbf{\Phi}^{*}$ is the conjugate of $\mathbf{\Phi}$. 

On the one hand, we can consider this categorization as a direct sum decomposition of the Lie algebra $\mathfrak{gl}(n)$. It is known from the literature~\cite{gilmore2012lie} that the sub-algebra of traceless matrices is the Lie algebra $\mathfrak{sl}(n)$ of the $SL(n)$ group (i.e., the special linear group). The anti-symmetric matrices form the Lie algebra $\mathfrak{so}(n)$ of the $SO(n)$ group (i.e., the special orthogonal group). 

On the other hand, we can interpret using elementary property from vector calculus that the symmetric, trace-free derivative operation physically relates to that of a \textit{shear}~\cite{romano2012no}. Mathematically this operation is represented by the matrix, $\mathbf{\Phi}_{1}$, which one can imagine as the gradient of a vector field in an arbitrary direction (usually expressed by a Jacobian matrix). However, the anti-symmetric matrix, $\mathbf{\Phi}_{2}$ represents pure rotation (i.e., the curl operator). 

At first, using the traceless symmetric matrix, $\mathbf{\Phi}_{1}$, we define a quantity as,
\begin{equation}
    P_{ns} = \frac{\sqrt{\dfrac{n}{n - 1}}{\left\lVert\mathbf{\Phi}_{1}\right\rVert}_{\text{F}}}{\tr(\mathbf{\Phi})}
    \label{eq:degree_shear}
\end{equation}
where ${\left\lVert \cdot \right\rVert}_{\text{F}}$ is the Frobenius norm of the matrix. We call this quantity the \textit{degree of shearing}. One can show that $P_{ns}$ is invariant under unitary transformation. In particular, $P_{3s}$ can be considered as the degree of polarization of the real part of the partially polarized $3 \times 3$ intrinsic coherency matrix, $Re(\mathbf{\Phi})$~\cite{Sheppard:22}. Moreover, it is interesting to note that we can relate the \textit{degree of shearing} to the component of polarimetric purity (CPP) proposed by Gil~\cite{gil2014interpretation,gil2016components} i.e., the \textit{degree of linear polarization}, $P_{l}$, for 2D and 3D cases, and the \textit{degree of directionality}, $P_{d}$, for the 3D case as,
\begin{align}
    P_{2s}^{2} &= P_{l}^{2} \label{eq:degree_shear_2d} \\
    P_{3s}^{2} &= \frac{3}{4}P_{l}^2 + \frac{1}{4}P_{d}^2 \label{eq:degree_shear_3d}
\end{align}
Thus, one can notice from equation~\eqref{eq:degree_shear_3d} that for the 3D case, the proposed \textit{degree of shearing} constitute not only the \textit{degree of linear polarization}, $P_{l}$, but also the \textit{degree of directionality}, $P_{d}$, which measure the stability of the plane that contains the polarization ellipse, or equivalently, the measure of closeness of the state represented by $\mathbf{\Phi}$ to that of a 2D state~\cite{gil2016components}. In a similar way, this can be further extended for $n > 3$. However, one needs an appropriate physical interpretation of such extension to higher dimensions. 

Further, using the anti-symmetric matrix, $\mathbf{\Phi}_{2}$, we express the \textit{degree of circular polarization} exactly as defined by Gil et al., as,
\begin{equation}
    P_{c} = \frac{\sqrt{2}{\left\lVert\mathbf{\Phi}_{2}\right\rVert}_{\text{F}}}{\tr(\mathbf{\Phi})}.
    \label{eq:degree_circular}
\end{equation}
This quantity measures all contributions to circular polarization and is also invariant under unitary transformation.

Finally, we express the \textit{generalized degree of polarimetric purity} by combining the \textit{degree of shearing}~\eqref{eq:degree_shear} and the \textit{degree of circular polarization}~\eqref{eq:degree_circular},  as,
\begin{equation}
    P_{nD} = \sqrt{P_{ns}^{2} + \frac{n}{2\,(n - 1)}P_{c}^{2}}.
    \label{eq:degree_purity}
\end{equation}
Furthermore, one can relate the \textit{generalized degree of polarimetric purity} for 2D and 3D cases to the three CPP parameters using equations~\eqref{eq:degree_shear_2d}, and~\eqref{eq:degree_shear_3d}, and equation~\eqref{eq:degree_circular} as,
\begin{align}
    P_{2D} &= \sqrt{P_{2s}^{2} + P_{c}^{2}} \\
    &= \sqrt{P_{l}^{2} + P_{c}^{2}} \label{eq:degree_purity_2d} 
\end{align}
and 
\begin{align}
    P_{3D} &= \sqrt{P_{3s}^{2} + \frac{3}{4}P_{c}^{2}} \\
    &= \sqrt{\frac{3}{4}P_{l}^2 + \frac{1}{4}P_{d}^2 + \frac{3}{4}P_{c}^2}. 
    \label{eq:degree_purity_3d}
\end{align}
In~\cite{gil2016components,gil2017polarized}, Gil has explicitly shown the relationships of $P_{2D}$ and $P_{3D}$ with the CPP parameters. Therefore, $P_{3s}$ provides fractional contributions from both $P_{l}$ and $P_{d}$ (i.e., polarization fraction due to mixed states), whereas $P_{2s}$ provides pure contribution from $P_{l}$. Using the derivation proposed in Sheppard et al.,~\cite{Sheppard:22}, we can show that,
\begin{align}
    P_{L}^{2} &= \frac{3}{4}P_{l}^{2} + \frac{1}{4}P_{d}^{2} - \frac{1}{4}P_{c}^{2},\\
    &= P_{3s}^{2} - \frac{1}{4}P_{c}^{2}
    \label{eq:sheppard_linear} 
\end{align}
where $P_{L}$ is defined in~\cite{Sheppard:22} as the degree of \textit{total linear polarization}, i.e., the contribution from both the purely polarized and mixed state. 

Now, expanding $P_{ns}$, and $P_{c}$ in the expression of $P_{nD}$ given in equation~\eqref{eq:degree_purity} in terms of the Frobenius norm and the matrix trace, we find that,  
\begin{align}
    P_{nD}^{2} & = \left(\dfrac{n}{n - 1}\right)\Bigg[\frac{{\left\lVert\mathbf{\Phi}_{1}\right\rVert}^{2}_{\text{F}} + {\left\lVert\mathbf{\Phi}_{2}\right\rVert}^{2}_{\text{F}}}{\left(\tr{\mathbf{\Phi}}\right)^{2}}\Bigg] \\
    & = \left(\dfrac{n}{n - 1}\right)\Bigg[\frac{ns^{2}}{n^{2}m^{2}}\Bigg], 
\end{align}
and therefore,
\begin{equation}
    P_{nD} = \frac{s}{\sqrt{n-1}\,m}.
\end{equation}
which is coincident with equation~\eqref{eqn:purity_bhattacharya}.

Hence, we suitably verify the equivalence among the two approaches for the expression of the \textit{generalized degree of polarimetric purity}, $P_{nD}$. 





\section{Acknowledgement}
The authors would like to thank Dr. David Roberts, the University of Adelaide, Australia, for the excellent blog article (\url{https://thehighergeometer.wordpress.com/2018/07/28/what-is-the-curl-of-a-vector-field-really/}, accessed on 09/06/2021) which prompted the initial conception of Approach~II of this work.
\bibliographystyle{IEEEtranN}
\bibliography{References}

\end{document}